
\sequentialequations

\def\prod#1#2{\langle #1 | #2 \rangle}

\def\ra{\rightarrow}

\def\frac#1#2{{#1 \over #2}}

\def\refmark#1{[#1]}

\PHYSREV

\frontpagetrue
{\baselineskip 14pt
\null
\line{\hfill FERMILAB-PUB-92/177-T}
\vskip 1.0cm

\centerline{\bf Vector-Boson versus Gluon Fusion at Hadron Colliders}

\vskip .75cm
\centerline{
{\bf J.~Bagger$^{(a)}$,}\foot{This work has been supported
by the U.S. National Science Foundation, grant
PHY-90-96198.}
{\bf  S.~Dawson$^{(b)}$}
\foot{This manuscript has been authored
under contract number DE-AC02-76-CH-00016 with the U.S.
Department of Energy.  Accordingly, the U.S. Government
retains a non-exclusive, royalty-free license to publish
or reproduce the published form of this contribution,
or allow others to do
so, for U.S. Government purposes.}
{\bf and G.~Valencia$^{(c)}$}}
\vskip .75cm
\centerline{$^{(a)}$ {\it Department of Physics and
Astronomy}}
\centerline{\it Johns Hopkins University, Baltimore,
MD~~21218}
\vskip 0.5cm
\centerline{$^{(b)}$ {\it Physics Department}}
\centerline{\it Brookhaven National Laboratory, Upton,
NY~~11973}
\vskip 0.5cm
\centerline{$^{(c)}$ {\it Theory Group}}
\centerline{\it Fermilab, Batavia, IL~~60510}
\vskip 1.75cm
\centerline{{\bf Abstract}}
\vskip 0.5cm

We use chiral perturbation theory to show that
pseudo-Goldstone boson scattering and gluon fusion
probe different aspects of electroweak
symmetry breaking at hadron colliders.
In particular, the physics responsible
for unitarizing the lowest-order pseudo-Goldstone boson
scattering amplitudes need not significantly affect the
gluon fusion process. We first show this
within the context of strict chiral perturbation theory,
and then discuss it using the language of resonances.

\vskip 1.25cm

\vfill
\line{June 1992 \hfill}}
\vfill
\eject

\overfullrule0pt

In a recent series of papers, we have used chiral
perturbation theory to study electroweak symmetry breaking at
high-energy hadron colliders
\REF\us{S.~Dawson and G.~Valencia, {\it Nucl. Phys.}
{\bf B352} (1991) 27;  J.~Bagger, S.~Dawson
and G.~Valencia, {\it Effective Field Theory Calculation of}
$pp \ra V_L V_L X$, Fermilab-Pub-92/75-T.}
\REF\usg{J.~Bagger, S.~Dawson and G.~Valencia,
{\it Phys. Rev. Lett.} {\bf 67} (1991) 2256.}
\REF\other{
T.~Appelquist and C.~Bernard, {\it Phys. Rev.} {\bf D22} (1980) 200;
A.~Longhitano, {\it Nucl. Phys.} {\bf B188} (1981) 118;
M.~Chanowitz and M.~Gaillard, {\it Nucl. Phys.} {\bf B261} (1985) 379;
M.~Chanowitz, H.~Georgi and M.~Golden, {\it Phys. Rev.} {\bf D36} (1987) 1490;
A.~Dobado and M.~Herrero, {\it Phys. Lett.} {\bf 228B} (1989) 495;
J.~Donoghue and C.~Ramirez, {\it Phys. Lett.} {\bf 234B} (1990) 361;
B.~Holdom, {\it Phys. Lett.} {\bf 258B} (1991) 156;
A.~Falk, M.~Luke and E.~Simmons, {\it Nucl. Phys.} {\bf B365} (1991) 523.}
\refmark{\us,\usg,\other}.
Our analysis is motivated by technicolor theories, in
which the Higgs sector of the standard model is replaced by
some sort of new, strongly-interacting physics that dynamically
breaks the electroweak symmetry.  The new physics induces
strong interactions between the would-be Goldstone bosons
that form the longitudinal components of the $W$ and $Z$. The interactions can
be parametrized quite generally in terms of
a low-energy effective Lagrangian.  The effective
Lagrangian describes the interactions of the would-be Goldstone
bosons at energies below the scale of symmetry breaking, and
below the masses of any other particles associated with the
symmetry breaking sector.

When the global symmetry group is larger than the usual
$SU(2)\times SU(2)$ of the standard model, the effective
Lagrangian contains new scalar fields, in addition to the usual
would-be Goldstone bosons.  In most models, these pseudo-Goldstone
bosons are relatively light, so they should be copiously produced
at the SSC or LHC.  In general, these particles affect the
scattering of longitudinal $W$'s and $Z$'s.  If some
carry color, as is typically the case, they also enhance the
gluon fusion production of longitudinal vector bosons.

To lowest order in an energy expansion, the effective Lagrangian
contains a single term:
$$
\eqalign{
{\cal L}^{(2)}\ =&\ {f^2 \over 4}\, {\rm Tr}\biggl(D_\mu \Sigma^\dagger
D^\mu \Sigma \biggr) \cr
\Sigma\ =&\  \exp \biggl({2i T^a \pi^a \over f} \biggr)\ ,\cr}
\eqn\lola
$$
where $f$ is the pseudo-Goldstone decay constant.
In this expression, the
$T^a$ are the (broken) generators associated with a global
symmetry group $G$, and ${\rm Tr} (T^a T^b) = {1\over2}\,\delta^{ab}$. For
$G = SU(2)\times SU(2)$, the fields $\pi^a$ are the would-be
Goldstone bosons associated with the $W$ and $Z$.  For
larger groups, they also include the pseudo-Goldstone bosons
discussed above.

The covariant derivative $D_\mu \Sigma$ describes the embedding of
$SU(3) \times SU(2)_L \times U(1)_Y$ into the symmetry group $G$.
For $G = SU(2N) \times SU(2N)$,  we find $f=v/\sqrt{N}$ in terms
of the usual $v \simeq 250$~GeV. The covariant derivative is given by
$$
D_\mu \Sigma\ =\
\partial_\mu \Sigma\
+\ {i\over2}\, g_s  G^\alpha_\mu  [T^\alpha,\Sigma]\
+ \ {i\over2}\, {g\over \sqrt N}  W^i_\mu T^i \Sigma \
-\ {i\over2}\, {g^{\prime}\over\sqrt N} B_\mu \Sigma T^3 \ ,
\eqn\covd
$$
where $G_\mu^\alpha$ is the $SU(3)$ color gauge field, and
$W_\mu^i$ and $B_\mu$ are the $SU(2)_L \times U(1)_Y$ gauge
bosons.  In this expression, the color coupling is written
as a commutator because color $SU(3)$ is contained in the
diagonal subgroup of $G$.  In contrast, $SU(2)_L$ is purely
left-handed, while $U(1)_Y$ acts on the right.  The
covariant derivative determines the lowest-order couplings
of the Goldstone particles to the standard-model gauge
bosons.

The lowest-order effective Lagrangian can be used to describe
pseudo-Goldstone boson scattering to order $p^2$ and gluon-gluon
scattering to order $p^4$ in the energy expansion.
To this order, the predictions
are universal, in the sense that they depend only on the
symmetry group $G$ and on the decay constant $f$.  At order $p^4$,
the pseudo-Goldstone scattering amplitudes depend on the
next-to-leading-order effective Lagrangian, which contains
four operators:
$$
\eqalign{
{\cal L}^{(4)}\ =&\ {L_1 \over 16 \pi^2}{\rm Tr}
\biggl(D_\mu \Sigma^\dagger D^\mu \Sigma \biggr)  {\rm Tr}
\biggl(D_\nu \Sigma^\dagger D^\nu \Sigma \biggr)\crr
& + \
{L_2 \over 16 \pi^2}{\rm Tr}
\biggl(D_\mu \Sigma^\dagger D^\nu \Sigma \biggr)  {\rm Tr}
\biggl(D_\mu \Sigma^\dagger D^\nu \Sigma \biggr) \crr
& +\ {N L_3 \over 16 \pi^2}{\rm Tr}
\biggl(D_\mu \Sigma^\dagger D^\mu \Sigma
D_\nu \Sigma^\dagger D^\nu \Sigma \biggr)\crr
& + \
{N L_4 \over 16 \pi^2}{\rm Tr}
\biggl(D_\mu \Sigma^\dagger D^\nu \Sigma
D_\mu \Sigma^\dagger D^\nu \Sigma \biggr)\ . \cr}
\eqn\nela
$$
The coefficients $L_i$ are expected to be of order one; they
are determined by the dynamics that underlie the symmetry
breaking.

It has been known for quite some time that the chiral
expansion breaks down at a scale of order $\Lambda \lsim 4
\pi v / N$ \
\Ref\scale{J.~Gasser and H.~Leutwyler {\it Ann. Phys.} {\bf 158} (1984) 142;
M.~Soldate and R.~Sundrum {\it Nucl. Phys.} {\bf B340} (1990) 1.}.
This value is based on the naive argument that loop
corrections must not dominate tree-level results.
Of course, the argument does not fix the scale very
precisely.  For the purist, this is irrelevant, since
the expansion only makes sense at energies much
lower than the scale of symmetry breaking. From a
practical point of view, however, one must decide
where to trust a given calculation.  It is often
suggested that the cutoff be placed at the scale where the
lowest-order pseudo-Goldstone boson scattering amplitudes
violate perturbative (two-body-elastic) unitarity,
a scale which is also of order $4\pi v/N$.

Since the lowest-order effective Lagrangian is unique, its
predictions do not distinguish between different dynamical
models of symmetry breaking.  These distinctions appear at
order $p^4$, in the form of the unknown coefficients
$L_i$.  For practical applications,
one would like to identify a region where
the energies are high enough for these terms to
be significant, and yet low enough for one to
trust the energy expansion.\foot{Unitarization prescriptions
have been proposed as a way to extend the region of validity
of the calculations \
\Ref\dob{See for example
A.~Dobado, M.~Herrero, and T.~Truong, {\it Phys. Lett.} {\bf 235B}
(1990) 129 and 134; R.~Willey, {\it Phys. Rev.} {\bf D44} (1991) 3646.}.
This is contrary to the spirit of effective Lagrangians,
where the fundamental physics is parametrized by the $L_i$.}

As a practical answer to this question, we have advocated
using the {\bf next-to-leading-order} chiral Lagrangian
in the vicinity of the scale where
the {\bf lowest-order} partial waves first violate
unitarity \Ref\snow{J.~Bagger, S.~Dawson and G.~Valencia, {\it Proceedings
of the 1990 Snowmass Study on Physics in the 90's}, Snowmass, CO.}.
The motivation for this choice is simple:

1. The energy is large enough for the $L_i$ to
be important.

2. The corrections stay within a factor of two of the lowest-order
amplitudes.

\noindent
Of course, we must now decide where to stop trusting
the next-to-leading results.  We assert that it is reasonable
to use the order $p^4$ chiral Lagrangian in the region where
the ${\cal O}(p^4)$ partial waves
preserve unitarity. Alternatively, we can also add
resonances to the model \
\Ref\reson{S.~Weinberg, {\it Phys. Rev.} {\bf 166}
(1968) 1568; S.~Coleman, J.~Wess and B.~Zumino, {\it Phys. Rev.}
{\bf 177} (1969) 2239; C.~Callan, S.~Coleman, J.~Wess and B.~Zumino,
{\it Phys. Rev.} {\bf 177} (1969) 2246; Applications to hadron
colliders can be found in J.~Bagger, T.~Han and R.~Rosenfeld,
{\it Proceedings
of the 1990 Snowmass Study on Physics in the 90's}, Snowmass, CO. and
references therein.} and preserve unitarity beyond the naive
counting scale $\Lambda$. In this case, the formalism should not be trusted
beyond the scale $\Lambda \sim 4\pi v/N$, about $800$~GeV for $N=4$.

Although chiral perturbation theory gives rigorous low-energy
results, it becomes more model-dependent when used at higher
energies.  We can still use it as a qualitative guide from which
we can infer general trends, but we should not take
any particular numbers very seriously.
If we wish to make detailed predictions for all energies, we must construct
(and solve) explicit models.  Given the large number of
possibilities  associated with electroweak symmetry breaking,
we believe that our more modest goals are both practical and reasonable.

Let us now compute the pseudo-Goldstone-boson scattering amplitudes
within this framework.  For $G = SU(2N) \times SU(2N)$,
spontaneously broken to $SU(2N)$, there are three channels
that grow with $N$ \
\Ref\casu{R.~Cahn and M.~Suzuki, {\it Phys. Rev. Lett} {\bf 67} (1991)
169.}.
They are the singlet, the symmetric adjoint and the antisymmetric
adjoint.  The $J=0$ partial wave in the singlet channel is
$$\eqalign{
{\rm Re}\ a_0^S\ =&\ {1\over 32 \pi} \biggl\{ {s n\over f^2}+
{s^2\over 2 \pi^2 f^4}\biggl[ L_1^r(\mu) \biggl(n^2-{1\over 3}\biggr)
 +L_2^r(\mu) \biggl(1+{n^2\over 3}\biggr) \crr
+&\ {L_3^r(\mu)\over 6} \biggl(4 n^2 - 5\biggr)
+ {L_4^r(\mu)\over 6} \biggl( 2 n^2 - 5\biggr)\biggr] \crr
-&\ {n^2 s^2 \over 3456  \pi^2 f^4}\biggl(150\log\biggl({s \over \mu^2}\biggr)
-11\biggr)\biggr\} \ ,\cr}
\eqn\sing
$$
while the $J=1$ partial wave in the antisymmetric adjoint channel is
$$
\eqalign{
a_1^{AA}\ =&\ {1 \over 192 \pi}\biggl[{ns \over f^2}-{s^2 \over 2 \pi^2 f^4}
\biggl(2L_1^r(\mu)-L_2^r(\mu) \crr
+&\ {n^2 \over 4}(L_3^r(\mu) - 2L_4^r(\mu))\biggr)
-{n^2 s^2 \over 576 \pi^2 f^4}\biggr]\cr}
\eqn\anti
$$
and the $J=0$ partial wave in the symmetric adjoint channel is
$$
\eqalign{
{\rm Re}\ a_0^{SA}\ =&\ {1 \over 64 \pi}
\biggl[{ns \over f^2}+{2s^2 \over 3 \pi^2 f^4}
\biggl(L_1^r(\mu)+2L_2^r(\mu) \crr
+&\ {n^2-5 \over 2} L_3^r(\mu) +{n^2-10 \over 4}L_4^r(\mu)\biggr) \crr
-&\ {n^2 s^2 \over 3456 \pi^2 f^4}\biggl( 60 \log \biggl( {s \over \mu^2}
\biggr)+1\biggr)\biggr]\ .\cr}
\eqn\symad
$$
In these formulae, $n= 2N$ and we have treated the pseudo-Goldstone
bosons as massless.
\REF\newbu{R.~S.~Chivukula, M.~J.~Dugan and M.~Golden, HUTP-92/A025,
BUHEP-92-18.}
Our amplitudes agree with those
of Ref.~\newbu, once we account for the fact that we
have chosen a different renormalization scheme:
$$
\eqalign{
L_1\ =&\ L_1^r(\mu)
-{1 \over 32}{1\over \hat{\epsilon}}-{1 \over 16} \cr
L_2\ =&\ L_2^r(\mu)
-{1 \over 16}{1\over \hat{\epsilon}}-{1 \over 8} \cr
L_3\ =&\ L_3^r(\mu)
-{1 \over 24}{1\over \hat{\epsilon}}-{5 \over 72} \cr
L_4\ =&\ L_4^r(\mu)
-{1 \over 48}{1\over \hat{\epsilon}}-{1 \over 18}\ . \cr}
\eqn\renp
$$
Our prescription is such that all constants that appear at order $p^4$
in the singlet, the antisymmetric adjoint and the symmetric adjoint
amplitudes are absorbed into the renormalized coefficients.

\FIG\pawaf{We show the partial wave amplitudes for
the singlet channel (a); the antisymmetric adjoint (b); and the
symmetric adjoint (c). The dotted curves represent the
real part of the lowest-order
results, and the dashed curves represent the real part of the order
$p^4$ results for $L_1^r(\mu)=-0.10$, $L_2^r(\mu)=0.30$,
$L_3^r(\mu)=-0.09$ and $L_4^r(\mu)=-0.13$,
all at $\mu=384$~GeV. The
solid line is the modulus of the amplitude
for a tree-level model with a vector resonance
of mass $400$~GeV and width $40$~GeV and a scalar resonance of
mass $400$~GeV and width $50$~GeV.}
In Figure~1 we have plotted these partial waves for some particular
values of the $L_i^r(\mu)$.  We see that there can be a considerable difference
between the energies at which the ${\cal O}(p^2)$
and the ${\cal O}(p^4)$ partial waves first
violate unitarity.
We associate the unitarity violation with the appearance of some
structure in the fundamental theory that cannot be properly
described by the low-energy constants (such as a resonance).\foot{If
there is such low energy structure, it is clear that it should
be studied directly and {\it not} by an energy expansion.} The
chiral Lagrangian is most useful when these structures are pushed
as high as possible, roughly, to the naive counting scale $\Lambda$.
For comparison we also present in this figure the result of a model
with vector and scalar resonances coupled via the Lagrangian
$$
{\cal L}\ =\ i{g_\rho f^2 \over  m_\rho}{\rm Tr}\biggl(\rho_{\mu\nu}
\xi^\dagger D_\mu \Sigma D_\nu \Sigma^\dagger \xi \biggr) +
{ g_\sigma f^2 \over m_\sigma}\sigma {\rm Tr}\biggl(D_\mu \Sigma D^\mu
\Sigma^\dagger\biggr)\ .
\eqn\lagres
$$
The couplings of the resonances to the pseudo-Goldstone bosons
can be written in terms of
the resonance widths.\foot{$\xi$ is such that $\xi\cdot \xi =\Sigma$.
The vector resonance
has been introduced as an antisymmetric tensor field following
Gasser and Leutwyler, Ref.~\scale.}

Let us next turn our attention to gluon fusion.  To lowest order
in chiral perturbation theory, ${\cal L}^{(2)}$ does not induce
the production of longitudinal gauge boson pairs by gluon fusion. This
process first occurs at order $p^4$ in the chiral expansion, that is,
at one-loop with ${\cal L}^{(2)}$ and tree-level with
${\cal L}^{(4)}$.  To this order, the gluon fusion
process is independent of the $L_i$. This is in sharp contrast
to the pseudo-Goldstone boson
scattering amplitudes, which depend on the $L_i$.
The amplitude for gluon fusion production of $Z_LZ_L$, for example, is
given by:
$$\eqalign{
M\bigg(g_\mu^\alpha (q_1) g^\beta_\nu (q_2) \ra& Z_L(p)
Z_L(p^\prime )\bigg)\crr
& = \
\epsilon^\mu (q_1) \epsilon^\nu (q_2)
\delta_{\alpha \beta }\sum_R T(R)
{\alpha_s \over \pi v^2} \biggl( -{s \over 2}g_{\mu \nu} + q_{2\mu}
q_{1\nu} \biggr)\ ,\cr}
\eqn\glufu
$$
where $T(R)$ is $1/2$ for each color triplet and $3$ for each color
octet.

For the $SU(8)$ Farhi-Susskind model, the lowest-order pseudo-Goldstone
boson scattering amplitudes violate unitarity
\Ref\buo{R.~S.~Chivukula, M.~Golden and  M.~V.~Ramana, {\it Phys. Rev. Lett.}
{\bf 68} (1992) 2883.}
at about 440 GeV.
To order $p^4$, however, this scale depends on the $L_i$.  For some choices of
the $L_i$, unitarity violation can be pushed near the naive
counting scale of $\Lambda \lsim 800$ GeV.
Therefore we claim that to order $p^4$,
it is reasonable to use Eq.~\glufu\ above threshold, up to about
800 GeV.  Above 440 GeV, the result becomes
more model-dependent,
and as we have said previously,
it is only intended to give a rough guide.

In a recent paper \refmark{\buo}, it has been argued that our
glue-glue amplitude
is an obvious overestimate of the physical amplitude because the
singlet $SU(8)$ partial wave violates unitarity above 440 GeV.
We wish to stress that this violation of unitarity occurs in
the pseudo-Goldstone boson channels and {\it not} in the gluon
fusion channel, and that the precise energy at which it occurs
is model-dependent.  As we have emphasized, to
order $p^4$ in chiral perturbation theory,
the glue-glue prediction is universal, while the scale of unitarity
violation in pseudo-Goldstone scattering is not.
To illustrate our point, we will first
discuss the related process $\gamma \gamma \ra \pi^0 \pi^0$.
We will then construct an explicit model that does {\it not}
violate unitarity below $\Lambda \sim 800$~GeV in the pseudo-Goldstone
boson scattering amplitudes, and yet still results in a gluon fusion production
of $Z_LZ_L$ pairs as large as the one we presented in Ref.~\usg.

\REF\cim{The Crystal Ball data are from, H.~Marsike et. al.,
{\it Phys. Rev.} {\bf D41} (1990) 3324.}
\REF\cimth{J.~Bijnens and F.~Cornet,
{\it Nucl. Phys.} {\bf B296} (1988) 557; J.~F.~Donoghue, B.~R.~Holstein,
and Y.~C.~Lin, {\it Phys. Rev.} {\bf D37} (1988) 2423; J.~Bijnens,
S.~Dawson and G.~Valencia, {\it Phys. Rev.} {\bf D44} (1991) 3555;
C.~Im, SLAC-Pub-5627; M.~R.~Pennington, DTP-92-32 (1992) and
references therein.}
\FIG\gammafig{The cross section for $\gamma \gamma \ra \pi^0
\pi^0$, from Ref.~\cim.  The solid line gives the ${\cal O}
(p^4)$ prediction in $SU(3)$ chiral perturbation theory.}

We show  in Figure~\gammafig~ the analogous QCD process
$\gamma \gamma \ra \pi^0 \pi^0$, and
the prediction from chiral perturbation
theory \refmark{\cim,\cimth}. We have presented the lowest-order
chiral perturbation theory result up to a very high energy, $1.4$~GeV,
where we have no reason to believe this prediction.
However, we see that
it does give a reasonable qualitative picture, accurate to a factor
of two, below the naive counting scale of 1.2 GeV.\foot{Recall that the
lowest-order $J=0$, $I=0$, partial wave amplitude in $\pi-\pi$ scattering
violates unitarity at about $500$~MeV.}

We now turn to our second and more important point, that the physics
of vector-boson scattering is very different from that of gluon fusion.
Within pure chiral
perturbation theory, this is clear when one goes to order $p^6$.
At this order, the gluon fusion process receives many contributions,
some of which do not directly affect pseudo-Goldstone-boson scattering.
For example, gluon fusion receives new contributions from the
operator:
$$
{\cal L}^{(6)}\ =\   {\alpha_s L \over 4 \pi \Lambda^2} {\rm Tr}
\biggl( D_\alpha \Sigma D^\alpha
\Sigma^\dagger\biggr) G^{a \mu \nu}
G^a_{\mu \nu} \ .
\eqn\psix
$$
These terms clearly do
{\it not} contribute to pseudo-Goldstone-boson scattering except at
one loop in QCD.  Given the size of $\alpha_s$, these contributions
can be safely neglected when compared to the potential
strongly-interacting electroweak symmetry breaking sector.

Unfortunately, at order $p^6$, chiral perturbation theory
is not practical due to the large number of operators that appear.
We can, however, resort to specific models; for example, including
resonances.
Above 440 GeV, it is possible for a resonance like the techni-rho
to unitarize the pseudo-Goldstone scattering amplitudes, as shown
in Figure~1.
However, the techni-rho does not appear as a resonance in
the gluon-fusion process. Although the techni-rho
certainly affects glue-glue
scattering, this mode is much more sensitive to resonances
which couple directly to the glue-glue intial state,
as a techni-sigma or techni-$f_2$  \
\Ref\fut{J.~Bagger, S.~Dawson, E.~Poppitz
and G.~Valencia, in preparation.}.  For example,
the following operator induces a
direct coupling between the techni-sigma and the gluons,
$$
{\cal L}\ =\ {\alpha_S h_\sigma  \over m_\sigma}
\sigma G^a_{\mu \nu} G^{a \mu \nu}\ .
\eqn\dirco
$$
\FIG\figlu{Production of $Z_L Z_L$ pairs by gluon fusion
at the SSC with a rapidity cut $|y|<2.5$. The dotted line
corresponds to one color octet of pseudo-Goldstone bosons.
The solid line corresponds to a techni-sigma of mass $400$~GeV,
width $50$~GeV and $h_\sigma =0.8$.}
In Figure~3 we show the contribution from a color octet of pseudos
to $Z_L Z_L$ production via gluon fusion at the SSC. We compare this
to the contribution from the techni-sigma of Eqs.~\lagres,~\dirco. We
have chosen the same values of mass and width used in Figure~1, as
well as $h_\sigma = 0.8$, a number of the same order as those
found in the model of Cahn and Suzuki \refmark{\casu}.
We see that in
this model, pseudo-Goldstone boson scattering does not violate
unitarity below $\Lambda \lsim 800$~GeV, and yet,
the production  of $Z_L Z_L$ pairs in gluon fusion is
comparable to that obtained in lowest-order chiral perturbation theory.

Direct couplings as that in Eq.~\dirco\ are completely unconstrained by the
unitarity of the pseudo-Goldstone scattering process.
For example, in QCD, the $f_2$ gives a major contribution
to photon-photon scattering, as shown in Figure~2.
At energies near 1 GeV, the resonant production of the $f_2$
provides the dominant contribution to the scattering
amplitude.  The only constraint on the
$f_2$ coupling to photons comes from
the $\gamma \gamma \ra \pi^0 \pi^0$ process itself.

In this letter we have argued that pseudo-Goldstone boson
scattering and gluon fusion probe different aspects of electroweak
symmetry breaking.  We have emphasized that different
physics couples to the two initial states.  In chiral
perturbation theory, this shows up in the different
operators that contribute to each process at a given
order in the energy expansion. In models with resonances
this shows up in the direct couplings of resonances to
the glue-glue state. We have performed our
analysis for chiral $SU(2N)\times SU(2N)$,
however, it is clear that
the same conclusions hold for the more usual case $N=1$. The
only difference being that $N=1$ does not allow additional
colored pseudo-Goldstone bosons. Although we have presented
results for massless pseudo-Goldstone bosons, we have checked
that including a small mass does not affect the qualitative
results of this paper.

\centerline{\bf Acknowledgements}

We are grateful to W.~Bardeen, E.~Eichten, C.~Im, C.~Quigg
and S.~Willenbrock for useful conversations.

\vfill
\eject

\refout
\figout
\bye